\documentclass{article}%
\usepackage{amsfonts}
\usepackage{amsmath}
\usepackage{amssymb}
\usepackage{graphicx}%
\newtheorem{theorem}{Theorem}

\newtheorem{definition}[theorem]{Definition}

\newtheorem{lemma}[theorem]{Lemma}

\newtheorem{remark}[theorem]{Remark}

\newcommand{\mj}{\mathrm{j}}
\begin{document}

\title{Zakharov-Shabat system and hyperbolic pseudoanalytic function theory}
\author{Viktor G. Kravchenko$^{\ast}$, Vladislav V. Kravchenko$^{\dagger}$ and
S\'{e}bastien Tremblay$^{\ddagger}$}

\date{}
\maketitle

\begin{center}
{$^{\ast}$Departamento de Matem{\'a}tica, F.C.T., Universidade do
Algarve,
\newline Campus de Gambelas, 8000-810 Faro, Portugal \newline$^{\dagger}%
$Departamento de Matem{\'a}tica, CINVESTAV del IPN, Unidad Quer{\'e}taro,
Libramiento Norponiente No.~2000 C.P. 76230 Fracc. Real de
Juriquilla,
Quer{\'e}taro, Mexico \newline$^{\ddagger}$D{\'e}partement de math{\'e}%
matiques et d'informatique, Universit{\'e} du Qu{\'e}bec,
Trois-Rivi\`eres, Qu\'ebec, G9A 5H7, Canada}
\end{center}

\begin{abstract}
In [1] a hyperbolic analogue of pseudoanalytic function theory was developed.
In the present contribution we show that one of the central objects of the
inverse problem method the Zakharov-Shabat system is closely related to a
hyperbolic Vekua equation for which among other results a generating sequence
and hence a complete system of formal powers can be constructed explicitly.
\end{abstract}

\section{Introduction}

In \cite{KRT}  a hyperbolic analogue of pseudoanalytic function theory was
developed using the algebra of hyperbolic numbers instead of complex numbers
in the case of Bers \cite{Bers52, AgmonBers, Bers56}. Hyperbolic numbers
$\mathbb{D}$, also called duplex numbers, is a commutative ring with zero
divisors defined in the plane by $\mathbb{D}={\large \{z=x+\mathrm{j}%
t:x,t\in\mathbb{R},\ \mathrm{j}^{2}=1\}}$ (see \cite{Sobczyk, MotterRosa,
GuoChunWen} for instance). With the aid of the hyperbolic pseudoanalytic
function theory in \cite{KRT} a procedure for constructing an infinite system
of solutions for the Klein-Gordon equation  ${\large (\square-\nu
(x,t))\varphi(x,t)=0}$ was introduced. This system is a hyperbolic analogue of
formal powers in the sense of Bers. In the present paper we consider the
Zakharov-Shabat system and show that it is closely related to a hyperbolic
Vekua equation. Using the above mentioned procedure an infinite system of its
solutions is obtained.

\section{Hyperbolic pseudoanalytic functions}

{\normalsize In this section we present some results of hyperbolic
pseudoanalytic  function theory. For more details see \cite{KRT}.}

{\normalsize We will consider the variable $z=x+t\mathrm{j}$, where $x$ and $%
t$ are real variables and the corresponding formal differential
operators
$$
\partial _{z}=\frac{1}{2}\left( {\partial _{x}+\mathrm{j}\partial _{t}}%
\right) \mbox{ and }\partial _{\bar{z}}=\frac{1}{2}\left( {\partial _{x}-%
\mathrm{j}\partial _{t}}\right) .
$$
Notation $f_{\bar{z}}$ or $f_{z}$ means the application of $\partial _{\bar{z%
}}$ or $\partial _{z}$ respectively to a hyperbolic function $f(z)=u(z)+v(z)%
\mathrm{j}$. We have the following result in the hyperbolic
function theory.

\begin{lemma}
{\normalsize Let $f(x+t\mathrm{j})=u(x,t)+v(x,t)\mathrm{j}$ be a
hyperbolic function where $u_x,u_t,v_x$ and $v_t$ exist, and are
continuous in a neighborhood of $z_0$. The derivative
$$
f^{\prime}(z_0 ) =\lim_{\overset{\scriptstyle z \rightarrow z_{0}}{%
\scriptscriptstyle (z-z_{0}\mbox{
}inv.)}}\frac{f(z)-f(z_{0})}{z-z_{0}}
$$
exists, if and only if
$$
f_{\bar{z}}(z_0)=0.  \label{CR2}
$$
Moreover, $f^{\prime}(z_0)=f_{z}(z_0)$ and $f^{\prime}(z_0)$ is
invertible if and only if $\det\mathcal{J}_{f}(z_0)\neq 0$.
\label{basic} }
\end{lemma}

{The hyperbolic pseudoanalytic function theory is based on
assigning the part played by $1$ and $\mathrm{j}$ in an arbitrary
hyperbolic function $f=u(x,t)1+v(x,t)\mathrm{j}$ to two
essentially arbitrary hyperbolic functions $F$ and $G$. We assume
that these functions are defined and twice continuously
differentiable in some open domain $\Omega \subset \mathbb{D}$. We
require that
$$
\mbox{Im}\{\overline{F(z)}G(z)\}\neq 0.  \label{vec01}
$$
Under this condition, $(F,G)$ will be called a ``generating pair''
in $\Omega $.
Notice that $\mbox{Im}\{\overline{F(z)}G(z)\}=\left\vert {%
\begin{array}{*{20}c}
{\mbox{Re}\{F(z)\}} & {\mbox{Re}\{G(z)\}}  \\
   {\mbox{Im}\{F(z)\}} & {\mbox{Im}\{G(z)\}}  \\
\end{array}}\right\vert .$ It follows, from Cramer's theorem, that for every
$z_{0}$ in $\Omega $ we can find unique constants $\lambda
_{0},\mu _{0}\in \mathbb{R}$ such that $w(z_{0})=\lambda
_{0}F(z_{0})+\mu _{0}G(z_{0})$. More generally we have the
following result. }

\begin{theorem}
{\normalsize Let $(F,G)$ be generating pair in some open domain
$\Omega $. If $w(z):\Omega \subset \mathbb{D}\rightarrow
\mathbb{D}$, then there exist
\textbf{unique} functions $\phi (z),\psi (z):\Omega \subset \mathbb{D}%
\rightarrow \mathbb{R}$ such that
$$
w(z)=\phi (z)F(z)+\psi (z)G(z),\mbox{ }\forall z\in \Omega .
$$
Moreover, we have the following explicit formulas for $\phi $ and
$\psi $:
$$
\phi (z)=\frac{\mathrm{Im}[\overline{w(z)}G(z)]}{\mathrm{Im}[\overline{F(z)}%
G(z)]}\mbox{, }\psi (z)=-\frac{\mathrm{Im}[\overline{w(z)}F(z)]}{\mathrm{Im}[%
\overline{F(z)}G(z)]}.
$$
\label{explicit} }
\end{theorem}

{\normalsize Consequently, every hyperbolic function $w$ defined
in some
subdomain of }$\Omega $ {\normalsize admits the unique representation $%
w=\phi F+\psi G$ where the functions $\phi $ and $\psi $ are real
valued. Thus, the pair $(F,G)$ generalizes the pair
$(1,\mathrm{j})$ which corresponds to hyperbolic analytic function
theory.  }

{\normalsize \smallskip \smallskip \hspace{0.5cm} We say that
$w:\Omega
\subset \mathbb{D}\rightarrow \mathbb{D}$ possesses at $z_{0}$ the $(F,G)$%
-derivative $\dot{w}(z_{0})$ if the (finite) limit
$$
\dot{w}(z_{0})=\lim_{\overset{\scriptstyle z\rightarrow z_{0}}{%
\scriptscriptstyle(z-z_{0}\mbox{ }inv.)}}\frac{{w(z)-\lambda
_{0}F(z)-\mu _{0}G(z)}}{{z-z_{0}}}  \label{lim1}
$$
exists. }

{\normalsize \hspace{0.5cm} The following expressions are called
the characteristic coefficients of the pair $(F,G)$:
$$
\begin{array}{ll}
\label{coefficients} a_{(F,G)} =-\displaystyle \frac{\bar{F}G_{\bar{z}}-F_{%
\bar{z}}\bar{G}}{F\overline{G}-\overline{F}G}, & b_{(F,G)} =\displaystyle%
\frac{FG_{\bar{z}}-F_{\bar{z}}G}{F\overline{G}-\overline{F}G}
\\*[2ex]
A_{(F,G)}=-\displaystyle\frac{\overline{F}G_{z}-F_{z}\overline{G}}{F%
\overline{G}-\overline{F}G}, & B_{(F,G)}=\displaystyle\frac{FG_{z}-F_{z}G}{F%
\overline{G}-\overline{F}G}.%
\end{array}%
$$
}

\begin{theorem}
{\normalsize Let (F,G) be a generating pair in some open domain
$\Omega $. Every hyperbolic function $w\in C^{1}(\Omega )$ admits
the unique
representation $w=\phi F+\psi G$ where $\phi ,\psi :\Omega \subset \mathbb{D}%
\rightarrow \mathbb{R}$. Moreover, the $(F,G)$-derivative $\dot{w}=%
\displaystyle\frac{\mathrm{d}_{(F,G)}w}{\mathrm{d}z}$ of $w(z)$
exists and has the form
\begin{equation}
\dot{w}=\phi _{z}F+\psi
_{z}G=w_{z}-A_{(F,G)}w-B_{(F,G)}\overline{w} \label{derivative}
\end{equation}%
if and only if
\begin{equation}
w_{\bar{z}}=a_{(F,G)}w+b_{(F,G)}\overline{w}.  \label{vekua}
\end{equation}%
}
\end{theorem}

{\normalsize \smallskip The equation (\ref{vekua}) can be
rewritten in the following form
\begin{equation}  \label{equivvekua}
\phi_{\bar{z}}F+\psi_{\bar{z}}G=0.
\end{equation}
Equation (\ref{vekua}) is called ``hyperbolic Vekua equation'' and
any continuously differentiable solutions of this equation are
called ``hyperbolic $(F,G)$-pseudoanalytic functions''. }

\begin{remark}
{\normalsize The functions $F$ and $G$ are hyperbolic
$(F,G)$-pseudoanalytic, and $\dot{F}\equiv \dot{G} \equiv 0$. }
\end{remark}

\begin{definition}
{\normalsize \label{DefSuccessor} Let $(F,G)$ and $(F_{1},G_{1})$
- be two
generating pairs in $\Omega$. $(F_{1},G_{1})$ is called \ successor of $%
(F,G) $ and $(F,G)$ is called predecessor of $(F_{1},G_{1})$ if%
\begin{equation*}
a_{(F_{1},G_{1})}=a_{(F,G)}\qquad\mbox{and}\qquad
b_{(F_{1},G_{1})}=-B_{(F,G)}.
\end{equation*}
}
\end{definition}

{\normalsize The importance of this definition becomes obvious
from the following statement. }

\begin{theorem}
{\normalsize \label{ThBersDer} Let $w$ be a hyperbolic
$(F,G)$-pseudoanalytic function and let $(F_{1},G_{1})$ be a
successor of $(F,G)$.
If $\dot{w}=W\in C^{1}(\Omega )$ then $W$ is a hyperbolic $(F_{1},G_{1})$%
-pseudoanalytic function. }
\end{theorem}

\begin{definition}
{\normalsize \label{DefAdjoint}Let $(F,G)$ be a generating pair.
Its adjoint
generating pair $(F,G)^{\ast}=(F^{\ast},G^{\ast})$ is defined by the formulas%
$$  \label{F*G*}
F^{\ast}=-\frac{2\overline{F}}{F\overline{G}-\overline{F}G},\qquad G^{\ast }=%
\frac{2\overline{G}}{F\overline{G}-\overline{F}G}.
$$
}
\end{definition}

{\normalsize The $(F,G)$-integral is defined as follows
$$
\int_{\Gamma }w\,\mathrm{d}_{(F,G)}z=F(z_{1})\mbox{Re}\int_{\Gamma
}G^{\ast }w\,\mathrm{d}z+G(z_{1})\mbox {Re}\int_{\Gamma }F^{\ast
}w\,\mathrm{d}z \label{integraldef}
$$
where $\Gamma $ is a rectifiable curve leading from $z_{0}$ to
$z_{1}$. }

{\normalsize If $w=\phi F+\psi G$ is a hyperbolic
$(F,G)$-pseudoanalytic function where $\phi $ and $\psi $ are real
valued functions then
\begin{equation}
\int_{z_{0}}^{z}\dot{w}\,\mathrm{d}_{(F,G)}\zeta =w(z)-\phi
(z_{0})F(z)-\psi (z_{0})G(z).  \label{FGAnt}
\end{equation}%
This integral is path-independent and represents the
$(F,G)$-antiderivative
of $\dot{w}$. The expression $\phi (z_{0})F(z)+\psi (z_{0})G(z)$ in (\ref%
{FGAnt}) can be seen as a ``pseudoanalytic constant'' of the
generating pair $(F,G)$ in  $\Omega $. }

{\normalsize A continuous function $W(z)$ defined in a domain
$\Omega $ will be called $(F,G)$-integrable if for every closed
curve $\Gamma $ situated in a simply connected subdomain of
$\Omega $ the following equality holds
$$
\oint_{\Gamma }W\mathrm{d}_{(F,G)}z=0.  \label{FGintegrable}
$$
}

\begin{theorem}
{\normalsize Let $W$ be a hyperbolic $(F,G)$-pseudoanalytic
function. Then $W $ is $(F,G)$-integrable. }
\end{theorem}

\begin{definition}
{\normalsize \label{DefSeq}A sequence of generating pairs $\big\{
(F_{m},G_{m})\big\}
$ with $m\in \mathbb{Z}$, is called a generating sequence if $%
(F_{m+1},G_{m+1})$ is a successor of $(F_{m},G_{m})$. If $%
(F_{0},G_{0})=(F,G) $, we say that $(F,G)$ is embedded in $\big\{%
(F_{m},G_{m})\big\}$. }
\end{definition}

{\normalsize %\begin{theorem}
%Let \ $(F,G)$ be a generating pair in $\Omega$. Let $\Omega_{1}$
%be a bounded domain, $\overline{\Omega}_{1}\subset\Omega$. Then
%$(F,G)$ can be embedded in a generating sequence in $\Omega_{1}$.
%\end{theorem}
}

\begin{definition}
{\normalsize A generating sequence $\big\{(F_{m},G_{m})\big\}$ is
said to
have period $\mu>0$ if $(F_{m+\mu},G_{m+\mu})$ is equivalent to $%
(F_{m},G_{m})$ that is their characteristic coefficients coincide.
}
\end{definition}

{\normalsize Let $w$ be a hyperbolic $(F,G)$-pseudoanalytic
function. Using a generating sequence in which $(F,G)$ is embedded
we can define the higher
derivatives of $w$ by the recursion formula%
\begin{equation*}
w^{[0]}=w;\qquad w^{[m+1]}=\frac{\mathrm{d}_{(F_{m},G_{m})}w^{[m]}}{\mathrm{d%
}z},\quad m=1,2,\ldots
\end{equation*}
}

\begin{definition}
{\normalsize \label{DefFormalPower}The formal power
$Z_{m}^{(0)}(a,z_{0};z)$ with center at $z_{0}\in \Omega $,
coefficient $a$ and exponent $0$ is defined as the linear
combination of the generators $F_{m}$, $G_{m}$ with real constant
coefficients $\lambda $, $\mu $ chosen so that $\lambda
F_{m}(z_{0})+\mu G_{m}(z_{0})=a$. The formal powers with exponents $%
n=1,2,\ldots $ are defined by the recursion formula%
\begin{equation}
Z_{m}^{(n)}(a,z_{0};z)=n\int_{z_{0}}^{z}Z_{m+1}^{(n-1)}(a,z_{0};\zeta )%
\mathrm{d}_{(F_{m},G_{m})}\zeta .  \label{recformula}
\end{equation}%
}
\end{definition}

{\normalsize This definition implies the following properties. }

\begin{enumerate}
\item[(i)] {\normalsize $Z_{m}^{(n)}(a,z_{0};z)$ is a $(F_{m},G_{m})$%
- hyperbolic pseudoanalytic function of $z$. }

\item[(ii)] {\normalsize If $a_1$ and $a_2$ are real constants,
then
$Z_{m}^{(n)}(a_1+\mathrm{j}a_2,z_{0};z)=a_1Z_{m}^{(n)}(1,z_{0};z)+a_2Z_{m}^{(n)}(\mathrm{j},z_{0};z).$
}

\item[(iii)] {\normalsize The formal powers satisfy the differential relations%
\begin{equation*}
\frac{\mathrm{d}_{(F_{m},G_{m})}Z_{m}^{(n)}(a,z_{0};z)}{dz}%
=nZ_{m+1}^{(n-1)}(a,z_{0};z).
\end{equation*}
}

\item[(iv)] {\normalsize The asymptotic formulas
\begin{equation*}
Z_{m}^{(n)}(a,z_{0};z)\sim a(z-z_{0})^{n},\quad z\rightarrow z_{0}
\end{equation*}
hold. }
\end{enumerate}

\section{Zakharov-Shabat system and a hyperbolic Vekua equation}
Inverse scattering problems involving coupling mode have been
investigated by many authors. When the medium concerned is treated
as a continuously varying one, the one-dimensional case is usually
associated with Zakharov-Shabat coupling-mode equations (see,
e.g., \cite{Lamb})
\begin{equation}
\partial_{x}n_{1}+ikn_{1}=s(x)n_{2},\qquad\partial_{x}n_{2}-ikn_{2}%
=-s(x)n_{1}, \label{ZS}%
\end{equation}
where the functions $n_1$, $n_2$ (the modes) and the potential
$s(x)$ are complex valued functions and the parameter $k$ (the
wave number) is complex. This
system is frequently considered as a Fourier transform of the following system%
\begin{equation}
\partial_{x}n_{+}+\partial_{t}n_{+}=s(x)n_{-},\qquad\partial_{x}n_{-}%
-\partial_{t}n_{-}=-s(x)n_{+}. \label{ZS1}%
\end{equation}

\bigskip Consider the following functions%
\[
u=n_{-}+n_{+},\qquad v=n_{-}-n_{+}.
\]
We have%
\[
\partial_{x}u-\partial_{t}v=sv,\qquad\partial_{x}v-\partial_{t}u=-su.
\]

\noindent This system can be written in the form
\begin{equation}
W_{\overline{z}}=-\frac{s(x)\mathrm{j}}{2}\overline{W} \label{VekZS}%
\end{equation}
where $z=x+\mathrm{j} t$ ($\mathrm{j}^{2}=1$), $W=u+\mathrm{j} v$,
$W_{\overline{z}}=\frac{1}{2}(\partial_{x}-\mathrm{j}\partial_{t})W$.

The coefficient in this hyperbolic Vekua equation in general is not
representable in the form of a logarithmic derivative of a scalar function
(see \cite{KRT}). Nevertheless we are able to construct a corresponding
generating pair:%
\[
F(x)=\cos S(x)-\mathrm{j} \sin S(x),\qquad G(x)=\sin S(x)+\mathrm{j}\cos
S(x),
\]
where $S$ is an antiderivative of $s$. Notice that $\operatorname*{Im}%
(\overline{F}G)\equiv1$.

In order to introduce the $(F,G)$-derivative in the sense of Bers let us
calculate the characteristic coefficients $A_{(F,G)}$ and $B_{(F,G)}$. For
this the following auxiliary formulae are helpful%
\[
F_{\overline{z}}=F_{z}=-\frac{s}{2}G\qquad\text{and}\qquad G_{\overline{z}%
}=G_{z}=\frac{s}{2}F.
\]
Then
\[
A_{(F,G)}=0\qquad\text{and}\qquad B_{(F,G)}=-\frac{s(x)\mathrm{j}}{2}%
\]
(we used the relations $F\overline{F}+G\overline{G}=0$ and $F^{2}+G^{2}=2$).
Thus, the $(F,G)$-derivative of solutions of (\ref{VekZS}) has the form%
\[
w=\overset{\cdot}{W}=W_{z}+\frac{s(x)\mathrm{j}}{2}\overline{W}%
\]
and is a solution of the equation%
\[
w_{\overline{z}}=\frac{s(x)\mathrm{j}}{2}\overline{w}%
\]
for which a generating pair can be constructed as well%
\[
F_{1}(x)=\cos S(x)+\mathrm{j}\sin S(x),\qquad G_{1}(x)=-\sin S(x)+\mathrm{j}%
\cos S(x).
\]
The generating sequence $\big\{(F_{m},G_{m})\big\}$ has then the form
\[
F_{m}=\cos S(x)+(-1)^{m+1}\mathrm{j}\sin S(x),\quad G_{m}=(-1)^{m}\sin
S(x)+\mathrm{j}\cos S(x),
\]
with
\[
\big(W^{[n]}\big)_{\overline{z}}=(-1)^{n+1}\frac{s(x)\mathrm{j}}{2}%
\overline{W^{[n]}}\ \ \Leftrightarrow\ \ W^{[n+1]}=\big(W^{[n]}\big)_{z}%
+(-1)^{n}\frac{s(x)\mathrm{j}}{2}\overline{W^{[n]}}.
\]
That is, it is periodic with a period $2$. In this case the whole system of
formal powers can be constructed explicitly. We find that
\[
F_{m}^{\ast}=G_{m},\quad \quad G_{m}^{\ast}=F_{m}.
\]

Let us now construct the formal powers of (\ref{VekZS}) on the
\emph{time-like} subdomain $\Omega=\{z=x+\mj t\ |\ 0<x<t<\infty$\}
of the hyperbolic plane. For $a,z_0\in \Omega$ with $a=a_1+\mj
a_2$ and $z_0=x_0+\mj t_0$, we have by definition
$Z^{(0)}(a,z_0;z_0)=\lambda F(z_0)+\mu G(z_0)=a_1+\mj a_2$, where
$\lambda,\mu\in \mathbb{R}$. We solve easily the system of two
linear equations and obtain $\lambda=a_1\alpha-a_2\beta$ and
$\mu=a_1\beta+a_2\alpha$, where we defined $\alpha=\cos S(x_0)$
and $\beta=\sin S(x_0)$. Therefore, we obtain
\[
Z^{(0)}(a,z_0;z)=(a_1\alpha-a_2\beta)F(z)+(a_1\beta+a_2\alpha)G(z).
\]
Let us now calculate $Z^{(1)}(a,z_0;z)$. For that we need to
calculate $Z^{(0)}_1(a,z_0;z)$ before. We have
$Z^{(0)}_1(a,z_0;z_0)=\lambda F_1(z_0)+\mu G_1(z_0)=a_1+\mj a_2$,
where $\lambda,\mu\in\mathbb{R}$. Again we solve easily the linear
system and find
$Z^{(0)}_1(a,z_0;z)=(a_1\alpha+a_2\beta)F_1(z)+(-a_1\beta+a_2\alpha)G_1(z)$.
Therefore, we obtain
\[
\begin{array}{rcl}
Z^{(1)}(a,z_0;z)&=&\displaystyle \int_{z_0}^z
Z^{(0)}_1(a,z_0;\zeta)\mathrm{d}_{(F,G)}\zeta \\*[2ex]
 &=& F(z)\Big[(a_1\alpha+a_2\beta)X^{(1)}+(a_1\beta-a_2\alpha)Y^{(1)}+(t-t_0)(-a_1\beta+a_2\alpha)\Big]\\*[2ex]
&+&
G(z)\Big[(-a_1\beta+a_2\alpha)X^{(1)}+(a_1\alpha+a_2\beta)Y^{(1)}+(t-t_0)(a_1\alpha+a_2\beta)\Big],
\end{array}
\]
where $X^{(1)}=X^{(1)}(x_0;x)$ and $Y^{(1)}=Y^{(1)}(x_0;x)$ are defined by
\begin{equation}
X^{(1)}(x_0;x):=\displaystyle \int_{x_0}^x \cos \big(2S(\xi)\big)\mathrm{d}\xi \quad \mbox{and} \quad Y^{(1)}(x_0;x):=\displaystyle \int_{x_0}^x \sin \big(2S(\xi)\big)\mathrm{d}\xi. \label{X1Y1}
\end{equation}
Now, if we want to find $Z^{(2)}(a,z_0;z)$ we need to
calculate first $Z^{(1)}_1(a,z_0;z)$; which is themselves obtained from $Z^{(0)}_2(a,z_0
;z)$. However, since the generating pairs are of period $2$ we have that $Z^{(0)}_2(a,z_0
;z)=Z^{(0)}(a,z_0;z)$. Then we obtain
\[
\begin{array}{rcl}
Z^{(1)}_1(a,z_0;z)&=&\displaystyle \int_{z_0}^z
Z^{(0)}(a,z_0;\zeta)\mathrm{d}_{(F_1,G_1)}\zeta \\*[2ex] &=&
F_1(z)\Big[(a_1\alpha-a_2\beta)X^{(1)}+(a_1\beta+a_2\alpha)Y^{(1)}+(t-t_0)(a_1\beta+a_2\alpha)\Big]\\*[2ex]
&&
+G_1(z)\Big[(a_1\beta+a_2\alpha)X^{(1)}+(-a_1\alpha+a_2\beta)Y^{(1)}+(t-t_0)(a_1\alpha-a_2\beta)\Big].
\end{array}
\]
We are now able to calculate $Z^{(2)}(a,z_0;z)$:
\[
\begin{array}{rcl}
Z^{(2)}(a,z_0;z)&=&2\displaystyle \int_{z_0}^z
Z^{(1)}_1(a,z_0;\zeta)\mathrm{d}_{(F,G)}\zeta \\*[2ex] &=&
F(z)\Big[(a_1\alpha-a_2\beta)X^{(2)}+(a_1\beta+a_2\alpha)\widetilde{X}^{(2)}+2(t-t_0)(a_1\beta+a_2\alpha)X^{(1)}\\*[2ex]
&&+(-a_1\beta-a_2\alpha)\widetilde{Y}^{(2)}+(a_1\alpha-a_2\beta)Y^{(2)}+2(t-t_0)(-a_1\alpha+a_2\beta)Y^{(1)}\\*[2ex]
&&+\frac{t-t_0}{x-x_0}(a_1\beta+a_2\alpha)I^{(2)}+\frac{t-t_0}{x-x_0}(-a_1\alpha+a_2\beta)\widetilde{I}^{(2)}+2(t-t_0)^2(a_1\alpha-a_2\beta)\Big]\\*[2ex]
&&
G(z)\Big[(a_1\beta+a_2\alpha)X^{(2)}+(-a_1\alpha+a_2\beta)\widetilde{X}^{(2)}+2(t-t_0)(a_1\alpha-a_2\beta)X^{(1)}\\*[2ex]
&&+(a_1\alpha-a_2\beta)\widetilde{Y}^{(2)}+(a_1\beta+a_2\alpha)Y^{(2)}+2(t-t_0)(a_1\beta+a_2\alpha)Y^{(1)}\\*[2ex]
&&+\frac{t-t_0}{x-x_0}(a_1\alpha-a_2\beta)I^{(2)}+\frac{t-t_0}{x-x_0}(a_1\beta+a_2\alpha)\widetilde{I}^{(2)}+2(t-t_0)^2(a_1\beta+a_2\alpha)\Big]
\end{array}
\]
where the functions $X^{(n)}$, $Y^{(n)}$, $\widetilde{X}^{(n)}$,
$\widetilde{Y}^{(n)}$, $I^{(n)}$ and $\widetilde{I}^{(n)}$,
depending on $x_0$ and $x$, are given by
\[
\begin{array}{l}
X^{(n)}(x_0;x):=n\displaystyle \int_{x_0}^x X^{(n-1)}(x_0;\xi)\
\cos \big(2S(\xi)\big) \mathrm{d}\xi,\
Y^{(n)}(x_0;x):=n\displaystyle \int_{x_0}^x Y^{(n-1)}(x_0;\xi)\
\sin \big(2S(\xi)\big)\mathrm{d}\xi, \\*[2ex]
\widetilde{X}^{(n)}(x_0;x):=n\displaystyle \int_{x_0}^x
Y^{(n-1)}(x_0;\xi)\ \cos \big(2S(\xi)\big) \mathrm{d}\xi,\
\widetilde{Y}^{(n)}(x_0;x):=n\displaystyle \int_{x_0}^x
X^{(n-1)}(x_0;\xi)\ \sin \big(2S(\xi)\big)\mathrm{d}\xi, \\*[2ex]
I^{(n)}(x_0;x):=n\displaystyle \int_{x_0}^x
X^{(n-1)}(x_0;\xi)\mathrm{d}\xi,\
\widetilde{I}^{(n)}(x_0;x):=n\displaystyle \int_{x_0}^x
Y^{(n-1)}(x_0;\xi)\mathrm{d}\xi,
\end{array}
\]
with $X^{(0)}(x_0,x)=Y^{(0)}(x_0,x)=\widetilde{X}^{(0)}(x_0,x)=\widetilde{Y}^{(0)}(x_0,x)=1$.

\section{Conclusion}
We have shown that the Zakharov-Shabat system is related to a
Vekua equation for which a generating sequence is found. Using the
generating sequence it is possible to obtain the associated formal
powers which are solutions of the given Vekua equation. Therefore,
we obtain an infinite set of solutions for the Zakharov-Shabat
system.

\subsection*{Acknowledgments}
The research of V. G. Kravchenko was partially supported by Centro
de An\'alise Funcional e Aplica\c{c}\~oes Aplica\c{c}{\~o}es do Instituto
Superior T\'ecnico (Portugal). V.~V. Kravchenko wishes to express
his gratitude to CONACYT  (Mexico) for supporting this work via
the research project 50424. The research of S.~Tremblay was
supported in part by grant from CRSNG of Canada.

\end{document}